\documentclass{kluwer}    

\begin{document}                                                                                   
\begin{article}
\begin{opening}         
\title{The Specific Frequency of Globular Clusters in Galaxies.}
\author{Bruce \surname{Elmegreen}}  
\runningauthor{Bruce Elmegreen}
\runningtitle{Globular Cluster Specific Frequency}
\institute{IBM Research Division, T.J. Watson Research Center\\
PO Box 218, Yorktown Hts. NY 10598 USA}
\date{September 17, 1999}

\begin{abstract} Variations in the number of globular clusters per unit
galaxy luminosity, which is the specific frequency, and the mass of
globular clusters per unit total available mass, are reviewed.
Correlations with galaxy luminosity and morphology, and with the
properties of the associated galactic clusters are discussed, as are
the color distributions of globular cluster populations. Several models
for the origin of these variations are summarized, including the tidal
stripping and interaction/starburst models. Most of the variations in
the specific frequency for elliptical galaxies result from variations in
the ratio of total mass to light with a nearly constant efficiency for
conversion of gas into globular cluster mass. Smaller additional
variations may come from tidal stripping, particularly in cD galaxies.

\end{abstract}

\end{opening}           

to be published in "Toward a New Millennium in Galaxy Morphology"
edited by D.L.  Block, I. Puerari, A. Stockton and D. Ferreira (Kluwer,
Dordrecht), 1999, from a conference in Johannesburg, South Africa,
September 13-18, 1999

\section{Introduction}  
Globular clusters (GC) were among the first objects to form in the Universe.
Their distributions relative to galaxies and clusters
of galaxies, their elemental abundances, motions, and internal
properties like radius and mass, all contribute to our understanding of
the origin of structure and the associated early star formation.

One of the fundamental parameters of globular cluster research has been
the specific frequency (Harris and van den Bergh 1981) \begin{equation}
S_N= N_{GC} 10^{0.4(M_V+15)}, \end{equation} which is the number of globular
clusters, $N_{GC}$, per unit galaxy luminosity, normalized to a galaxy with an
absolute V magnitude of $-15$. Spiral galaxies typically have $S_N\le1$
(van den Bergh and Harris 1982), while dwarf ellipticals (Durrell et al.
1996; Miller et al. 1998), ellipticals (Harris and van den Bergh 1981),
and S0 galaxies (Kundu and Whitmore 1998; Chapelon et al. 1999) have
$S_N$ in the range from 2 to 6, with lower values in more isolated
regions (see review in Harris 1991). Giant cD galaxies near the centers
of large clusters have larger $S_N\sim10-20$ (Harris and van den Bergh
1981; Harris, Pritchet and McClure 1995; Blakeslee and Tonry 1995; Bridges
et al. 1996), with some exceptions (e.g., Kaisler et al. 1996).

This variation of $S_N$ from galaxy to galaxy, particularly along
morphological classes, has been a long-standing problem in globular
cluster research. It seems appropriate at a morphology 
conference like this to
review the nature of this problem and some of the solutions that have
been suggested. Unfortunately, there is no completely satisfactory
explanation that accounts for all of the available information about
globular cluster distributions and properties, but perhaps some
combination of the suggested solutions is appropriate. In any case, the
observations tell us a lot about the formation of globular clusters,
galaxies, and clusters of galaxies. 

\section{Properties of globular clusters related to the Specific Frequency}

\subsection{$S_N$ Correlations with Galaxy Luminosity}

\subsubsection{Observations}

Djorgovski and Santiago (1992) found that $S_N$ increases with the
luminosity $L$ of the host galaxy, meaning that the actual number of 
GC's increases faster than the first power of $L$.  Zepf, Geisler,
and Ashman (1994) confirmed this using data from the literature, after
first noting that a bright non-cluster elliptical, NGC 3923, has a
moderately high $S_N$. 

The question of whether the increase in $S_N$ with $L$ is continuous or
step-wise was investigated by Kissler-Patig (1997), who distinguished
between faint and bright systems: Faint ellipticals are disky with
unresolved cores, and they have low $S_N$ and low metallicities. Bright
ellipticals are boxy with resolved cores, and they have high $S_N>5$ and
bimodal distributions of GC colors (Sect. \ref{sect:bimodal}).

At low galaxy luminosity, particularly among nucleated dwarf
ellipticals, the specific frequency decreases with increasing $L$, which
is opposite from the case for high $L$ galaxies; non-nucleated dE's have
no $L$ trend at all (Miller et al. 1998). However, the average $S_N$ for
dE's is about the same as for bright ellipticals (Durrell et al. 1996),
so the decrease in $S_N$ with decreasing $L$ for normal ellipticals
turns around at low enough $L$ for the dwarf ellipticals.

\subsubsection{Theory}

The origin of the $S_N(L)$ relation has had various explanations. 
Aside from the extremely large values of $S_N$ for cD galaxies, 
some of the $S_N$ variations among ellipticals is
from variations in the mass-to-light ratio, which is known to 
increase with $L$ (Faber et al. 1987; van der Marel
1991; Pahre et al. 1995).  

Kissler-Patig et al. (1997) note that the
number of globular clusters in spiral galaxies per unit luminosity
of the spheroid, rather than the whole galaxy, is about the same
as the number per unit total luminosity for the ellipticals. 
They suggest that it makes more sense to compare the spheroids anyway
because the disks of spirals are still bright with active star formation.

The decrease in $S_N$ with increasing $L$ for dE galaxies has been
explained as a result of gas removal during a starburst phase in the
main galaxy, after the globular clusters formed (Durrell et al. 1996).
Gas removal lowers the mass and luminosity of the galaxy today, but does
not change the number of globular clusters. Thus the number per unit
luminosity increases with time. If gas removal is more important for
lower mass systems, then this increase in $S_N$ from birth 
would be larger for them, explaining the observed correlation. 

McLaughlin (1999) explains the whole $S_N(L)$ correlation from dE's to
giant ellipticals, including cD's, as a combination of effects that
operate together while the number of globular clusters $N_{GC}$ per unit
total mass remains constant. These effects are: (1) an increase in the
star mass-to-light ratio with $L$, noted above; (2) an increase in the
x-ray gas mass with $L$, and
(3) the selective self-destruction of low mass dE galaxies, discussed
above. Items (1) and (2) give the observed increase in
$S_N(L)$ for elliptical galaxies at constant $N_{GC}$/total mass.
Item (3) considers the same {\it initial} ratio of the number of
clusters to the total mass. 

The step-wise distinction made by Kissler-Patig (1997) was similarly
explained by McLaughlin (1999) after noting that boxy ellipticals have
higher x-ray emitting masses than disky ellipticals (Bender et al.
1989). If specific frequency is measured per unit total mass, rather
than V-band luminosity, then the two-step distinction of globular
cluster populations vanishes. 

One of the most compelling early explanations for the increase in $S_N$
with $L$ and the variation with Hubble type is that spiral galaxies
combine to make ellipticals, and the resulting starbursts during the
mergers make large numbers of globular clusters (Ashman and Zepf 1992).
There are several problems with the details of this scenario, as
discussed in Section \ref{sect:intmodel}, but some elliptical galaxies
probably {\it do} form by interactions (Toomre 1977, and see below), and
globular clusters {\it do} form in great quantities in interacting
galaxies, possibly with large enough numbers to increase $S_N$ from the
spiral-galaxy value to a low elliptical galaxy value (Schweizer et al.
1996). Interaction models are also interesting because of their
prediction of bimodal color distributions for GC populations (Sect.
\ref{sect:bimodal}). However, most studies of GC populations in old
systems discount the standard interaction scenario in which spirals
commonly merge to make ellipticals while increasing $S_N$. This will be
discussed in more detail below. 

Another explanation for the $S_N(L)$ correlation considers the continuous
destruction of GC systems long after they form. Murali and Weinberg
(1997) reproduced the observed $S_N(L)$ with a variable destruction rate
following a uniform $S_N$ at birth. The destruction rate varies with the
luminosity of the host because the density of elliptical galaxies varies
with $L$. Higher density ellipticals, which have lower $L$, destroy
their globular clusters faster, giving these ellipticals lower $S_N$
today. Murali and Weinberg (1997) also predicted that the mass at the
peak of the GC luminosity function should be larger for more completely
destroyed GC systems, i.e., for lower mass, lower $S_N$, ellipticals.
This is because most GC destruction processes, particularly evaporation
by 2-body encounters, operate faster for lower cluster masses. 

However, the {\it GC luminosity function is not observed to change much}
from galaxy to galaxy (e.g., Blakeslee, Tonry and Metzger 1997; Kundu
and Whitmore 1998), or with position in a galaxy or cluster of galaxies
(Harris and Pudritz 1994; Harris, Harris and McLaughlin 1998; Kundu et
al. 1999), so the effects of such destruction would have to be confined
to unobservably low GC masses, making it inconsequential for $S_N$.
Nevertheless, the issue of GC destruction can be as important as GC
formation in producing $S_N$ and other GC correlations, as in the
classic explanation for the GC luminosity function by Fall and Rees
(1977). More studies of the disruption processes are clearly warranted.

\subsection{$S_N$ Correlations with Hubble Type and Morphology}

There are two obvious correlations between $S_N$ and galaxy morphology:
a sharp increase in $S_N$, by a factor of $\sim6$, going from spirals
where $S_N\le1$ to S0 and elliptical galaxies (van den Bergh 1984; van
den Bergh 1995a), and another sharp increase, by about a factor of
$\sim5$, going from bright cluster ellipticals to cD galaxies. 

The spiral value is low partly because spirals still have a lot of star
formation. If the spiral luminosity is corrected for fading and
extinction, then $S_N$ comes to within a factor of $\sim3$ of
the value for ellipticals (van den Bergh 1995a). Or, if the spiral value
is measured in terms of the mass of the spheroidal component only, then
it becomes comparable to the elliptical value (Kissler-Patig et al.
1997). 

The second jump to high $S_N$ for cD galaxies has had several
explanations. Most of these would seem to be of historical interest only
because McLaughlin (1999) found that cD galaxies have normal globular
cluster numbers per unit {\it total mass} when the x-ray emitting gas
mass is considered. Thus the oddity about cD galaxies may be only that
they have three times more optically {\it hidden} mass than other
ellipticals in the form of dark matter (high $M/L$ values) and x-ray
gas.  In what follows we review two additional models that have been proposed
to increase $S_N$: tidal stripping of globular clusters from smaller
galaxies, and interaction-induced globular cluster formation.

\subsubsection{Tidal Stripping Model for Excess Globular Cluster Counts in 
cD Galaxies}

There are several interesting explanations for the excess GC count in cD
galaxies. One of the most compelling is the stripping scenario.
Introduced by Forte et al. (1982) after a suggestion by van den Bergh
(1977), and simulated in some detail by Muzzio (1987) and Cote et al.
(1998), this model proposes that giant cD galaxies in the centers of
rich clusters have selectively stripped globular clusters from the
outskirts of neighboring galaxies, thereby increasing $S_N$ for the cD
and decreasing it for the neighbors. 

Detailed observations of the supposedly stripped neighbors of some cD
galaxies make this scenario plausible. Forbes et al. (1997) suggested
that if NGC 4486B near the cD galaxy M87 in Virgo originally had a
luminosity given by its velocity dispersion, and if it had a metallicity
given by this original luminosity, using the standard correlations, then
this companion galaxy must have lost 95\% of its stars since birth and
had a metallicity in the middle of the range for the metallicities of
the globular clusters in M87. In terms of numbers of GC transferred,
95\% of the original GC count for NGC 4486B would have been $\sim1000$
GC, and this is one-fifth of the total in M87 that have the same
metallicity. 

The same calculation can be done for NGC 1404, near the cD NGC 1399 in
Fornax (Forbes et al. 1997). If the observed number of GC in NGC 1404
today is subtracted from the expected number of GC at birth, given an
initial ratio $S_N=5$, then about one-third of the GC in NGC 1399 at the
same metallicity could have been transferred from NGC 1404. However, NGC
1404 has a luminosity today that is the same as what would be expected
from its velocity dispersion. This implies that NGC 1404
could not have lost a large fraction
of its stars.  Yet the difference in GC counts from a fiducial value with
$S_N=5$ represents a
50\% loss. Perhaps the transferred GC were in the outer parts of
NGC 1404, so only the GC went without any accompanying stars. In any
case, the tidal radius expected for NGC 1404 is about the same as the
observed radius (Forbes et al. 1997), so it looks tidally truncated.
NGC 1404 is also located in the x-ray halo of NGC 1399. 

Also for NGC 5846A near the larger galaxy NGC 5846 (Forbes et al. 1997),
the original metallicity of the smaller galaxy matches the metal poor
component of the GC population in NGC 5846, and the number of GC's in the
original NGC 5846A ($\sim800$) matches the current number at this
metallicity in NGC 5846. 

Neilsen et al. (1997) made a similar study of NGC 4478, another
companion to M87. The GC population in NGC 4478 has only one metallicity
value, and this is the same as for the metal-poor population in the
two-component color distribution in M87. NGC 4478 also has very few GC
today, giving it $S_N=0.6$ instead of the usual $S_N\sim5$ for this type
of galaxy. As a result, it might have lost $\sim300$ clusters. NGC 4478
is also so close to M87 that it is currently projected inside the
extended GC halo of the cD galaxy. Thus $\sim300$ of the blue GC in M87
could have come from NGC 4478. 

Also in support of the stripping model is that the velocity dispersion
of the GC system around the cD galaxy NGC 1399 is more similar to the
high velocity dispersion of the whole galactic cluster than the smaller
dispersion of NGC 1399 itself (Grillmair et al. 1994; Minniti et al
1998; Kissler-Patig et al. 1999). The same is true for cD M87 (Mould et
al. 1987; see Sect. \ref{sect:extra}). Thus many of the GC's in NGC 1399
and M87 could have had an origin outside the galactic boundary. The
specific frequency for ellipticals in general also seems to correlate
better with the galactic cluster velocity dispersion than with the host
galaxy dispersion (Blakeslee et al. 1997; Harris et al. 1998;
Kissler-Patig et al. 1999). Specific frequency correlates with local
galaxy density as well (Kumai et al. 1993a; West 1993; Blakeslee 1997;
Blakeslee et al. 1997).

In addition, the distribution of GC density with 
galactocentric distance is steeper in the companion galaxies that are 
supposedly stripped than in the cD, which received the
stripped GC (Forbes et al. 1997), so the stripped galaxies look
tidally truncated in their GC distributions.
Furthermore, the distribution of GC's in presumably unstripped dE galaxies is
more extended than the galaxy light (Minniti et al. 1996), so 
preferential stripping of the GC's alone, which is needed to
increase $S_N$ in the central galaxy, would seem to be possible. 

The problem with stripping models is that they are not likely to 
increase $S_N$ by the necessary amount, which is a factor of $\sim3$,
to explain the high values of this ratio in cD galaxies
(West 1993; Harris et al. 1998).  The accumulation of dwarf
galaxies by a cD will not increase $S_N$ much either because
dwarfs have relatively low $S_N\sim4-6$ themselves (Durrell et al. 1996;
Minniti et al. 1996).
Even if the extended GC envelopes of dE galaxies were alone
stripped to make a high $S_N$ in a cD galaxy, the remaining dE cores
are not obviously present nearby (Harris et al. 1998). 

Hilker et al. (1999) point out, however, that if the lowest mass
dwarfs have values of $S_N$ as high as the Fornax and Sagittarius dwarf
spheroidals ($S_N\sim20-30$), or, at least, if half the
dwarfs with magnitudes
fainter than $-12.5$ have one or more GC's, and if the residual gas in
the dwarfs also makes GC's as in interacting starburst systems, then
dwarf galaxy accretion by cD galaxies can provide the right number of
globular clusters to account for an $S_N\sim10$.  In addition,
they show that the surface density of dwarf galaxies drops below the
extrapolated lower-law distribution toward the center of the Fornax
cluster of galaxies, at the same place where the blue globular clusters
and cD halo stars around NGC 1399 have an overabundance.

\subsubsection{Interaction Model for the excess $S_N$ in Ellipticals}
\label{sect:intmodel}

The difference in $S_N$ between spiral and elliptical galaxies, which
is a factor of $\sim3-6$ depending on how it is counted, 
was long used as an argument against
the model in which ellipticals formed from the merger of spirals
(van den Bergh 1984).  Then Schweizer (1987) and 
Ashman and Zepf (1992) suggested
that interactions in the early Universe between gaseous spirals
made new GC's, thereby increasing $S_N$ to the required value by the
time the merged remnants looked like ellipticals.  Ashman
and Zepf predicted
that ellipticals should have two populations of GC's, one metal
poor from the original spiral GC population, and another metal
rich, from the more recent epoch of GC formation during the merger.
Remarkably, such bimodal GC populations were soon found
(Zepf and Ashman 1993), as discussed in more detail in 
section \ref{sect:bimodal}. 

The merger model for the formation of GC's was also confirmed directly by
the observation of profuse GC formation during recent gassy mergers,
some of which also end up looking like elliptical galaxies (Schweizer 1987;
Schweizer et al. 1996; Miller et al. 1997). Starbursts
in general make a lot of GC-like objects (Holtzman et al. 1992; Whitmore
et al. 1993; Whitmore and Schweizer 1995; see review in Ho 1997). 

Other predictions of the merger model have not faired so well, however,
and it is unclear today how much mergers played a role in
determining the range of $S_N$. 

One early problem was that mergers were expected to form 
a lot of stars that are not
in globular clusters, so it was unclear why the merger starburst
should have the required high $S_N$ by itself, 
and not a ``normal $S_N$'' from normal star formation (West 1993).
This problem may have gone away, however, because starburst
mergers do seem to put an unusually high fraction
of their young stars into GC's rather than unbound associations
(Meurer et al. 1995; Schweizer et al. 1996).

Another problem is that S0 galaxies have the same high $S_N$
as elliptical galaxies, and no one expects S0 galaxies to 
have been made by major mergers (West 1993). 

Also, the lowest metallicities for GC systems in elliptical
galaxies are higher than the normal metallicities of GC's in spiral
galaxies, so the old population of GC's in ellipticals
does not look like a GC population from a former spiral
(Geisler et al. 1996). 

Forbes et al. (1997) listed several other arguments against the
merger scenario: (1) metal-poor GC's in elliptical galaxies
are usually more numerous than metal-rich GC's, and high $S_N$
galaxies have more metal-poor GC's, so the excess in $S_N$ cannot
be from the {\it recent} formation of GC's at high metallicity
in a merger. (2) High $S_N$ galaxies do not have their metal-rich 
GC's concentrated in the center where the old starburst
should have occurred; rather they have their metal-rich
GC's distributed throughout the galaxy, with a large core radius.
(There is a slight red GC concentration, however.)
(3) A typical cD galaxy needs to have acquired $\sim10$ L$^*$
spirals to get its large mass, and these were likely to have
entered at different times in the past, producing different
metallicity values for the young GC components; however,
cD's generally have only one metal rich component of GC's; 
(4) some elliptical galaxies with the same luminosity have very different
$S_N$, so the mergers would have had to produce GC's with different
efficiencies; and (5) the main, centralized part of a cD galaxy
is kinematically distinct from its extended envelope, such that the
velocity dispersion of the envelope is more like the value in the whole galaxy
cluster than the value in the cD core; in the merger scenario, both the
core and the envelope should have been formed by the same mergers.

Kissler-Patig et al. (1998a) and Neilsen and Tsvetanov (1999) seconded
the point (1) above. In addition, the metal rich component of GC
populations has a metal abundance that scales with the host galaxy
luminosity, suggesting it is primordial rather than triggered by
interactions (Forbes et al. 1997). When there is a high
proportion of high-metallicity GC's in a galaxy, the $S_N$ ratio tends
to be low, so large $S_N$ values should be identified with the
primordial metal-poor component of GC's, rather than a younger component
that is metal-rich (Forbes et al. 1997). Harris (1995) and Harris et al.
(1998) also argue against the merger model for large $S_N$ galaxies, but
say that spiral mergers may account for the low values of $S_N\sim2$ in
relatively isolated ellipticals. 

Geisler et al. (1996) gives a specific example where
the merger model seems to fail: the two metallicity
peaks in the elliptical galaxies M49 and M87 have the same 
values, and the same ratios of high to low metallicity GC's, 
but $S_N$ is much higher in M87 than in M49. Thus, the excess in
$S_N$ for M87 cannot be related to its young, metal rich
population.  Also, the giant elliptical/S0 M86 has only the blue
GC population (Neilsen and Tsvetanov 1999).

\subsubsection{Summary of $S_N$ dependence on Morphology}

After McLaughlin (1999) showed that $S_N$ variations with
morphology can be explained mostly by variations in the
mass-to-light ratio and the amount of x-ray gas, so that
the {\it mass} ratio of GC mass to total baryon mass is about constant
even though the {\it luminosity} ratio in the definition of
$S_N$ is not, there is little compelling reason to look 
elsewhere for effects that increase $S_N$ substantially.

Nevertheless, gas-rich galaxy mergers may have increased $S_N$
a little (to $S_N\sim2$) in some ellipticals, and cD galaxies
may have acquired large numbers, but perhaps only modest 
fractions, of their
GC's from stripped neighbors.

\section{Bimodal Color Distributions for Globular Cluster Systems}
\label{sect:bimodal}

An important clue to the early history of elliptical galaxies
is the color distribution, 
or equivalently, metallicity distribution of
their globular clusters. Because of line blanketing at short wavelengths in 
metal-rich stars, red old clusters have more metals than blue old 
clusters. 

A remarkable observation is that globular cluster populations in
giant elliptical galaxies tend to have two components, red and blue
(Harris et al. 1992; Ostrov et al. 1993;
Zepf and Ashman 1993; Lee and Geisler 1993; Secker et al. 1995; 
Zepf, Ashman and Geisler 1995).

Geisler et al. (1996) studied M49, and found that the metal-rich GC
component is more concentrated in the center of the galaxy and the GC's
are more metal poor than the galaxy at all radii. Geisler et al. also
compared the GC populations in 5 other giant ellipticals previously
studied and found that the metallicity peaks are always separated by
about 1 dex, and that the most metal poor GC system is always more metal rich
than the GC's in spiral galaxies. This latter point was also made by
Forbes et al. (1996) for 14 elliptical galaxies, although Kissler-Patig
et al. (1998a) found that the most metal poor GC's in the
cD galaxy NGC 1399 have the same metallicity as the halo GC's in the Milky
Way. 

The metal-rich component of GC's seems to be intimately connected with
the host galaxy because the metallicity of this component scales with
the galaxy luminosity whereas the metallicity of the metal poor
component does not (Forbes et al. 1997; Cote et al. 1998). Also, 
in M49 at least, the metallicity of the metal-rich GC component nearly
mimics the metallicity of the galaxy halo (Geisler et al. 1996);
in M87, the metal-rich component is only slightly more metal poor
than the galaxy (Kundu et al. 1999). In
the S0 galaxy NGC 3115, the metal-rich GC component follows the thick
disk light distribution, whereas the metal poor component follows the
halo (Kundu and Whitmore 1998).

Only the brightest elliptical galaxies, including some cD's, have
bimodal color and metallicity distributions for the GC
systems (Kissler-Patig 1997).  Yet surprisingly, the bright
elliptical M86 in Virgo, does not; it has only the blue,
metal-poor component (Neilsen and Tsvetanov 1999).  M86 is also 
bluer than the other ellipticals as a whole, so presumably what
causes the red GC populations in other galaxies also gives their hosts
the same high metallicity. 

Another oddity is that the red GC radii tend to be 20\% smaller than the
blue GC radii in the two galaxies where this has been measured:
NGC 3115 (Kundu and Whitmore 1998) 
and M87 (Kundu et al. 1999).

The bimodal nature of GC systems in most giant elliptical galaxies is
not understood. There were clearly several epochs of GC formation
involved, but whether they took place inside these galaxies in two
bursts of star formation, or outside the galaxies in separate systems
that later merged, is not known. The connections between either
component, galaxy interactions, and $S_N$ are not clear either. For
example, Geisler et al. (1996) noted that M49 and M87 have the same
bimodal color distributions for their GC systems, but M87 has much
higher $S_N$ than M49. 

\section{Inter- and Extra-galactic formations of globular clusters}
\label{sect:extra}

The GC systems in the outer parts of cD galaxies, particularly the
blue or metal-poor components, tend to correlate in 
their properties with the whole galactic cluster rather than the
galaxy itself. 

The various galaxy-cluster correlations are of three types:
\begin{itemize}

\item {\it cluster density}: $S_N$ increases with the local
density of galaxies in a cluster
(Kumai et al. 1993a; West 1993; Blakeslee 1997; Blakeslee et al. 1997).

\item {\it cluster velocity dispersions}: 
the velocity dispersions of GC systems near cD galaxies are much larger
than the dispersions of the galaxy stars and more comparable to the
dispersion in the cluster as a whole, often
increasing continuously with radius in the cluster 
(Mould et al. 1987; Huchra and Brodie 1987;
Mould et al. 1990; Brodie and Huchra 1991; Grillmair et al. 1994;
Cohen and Ryzhov 1997; Minniti et al. 1998; Kissler-Patig et al. 1998a,
1999).
Also, $S_N$  
increases with the velocity dispersion of the {\it host} galaxy
(Kumai et al. 1993a) and with the velocity dispersion of the whole
cluster (Blakeslee 1997; Blakeslee et al. 1997; Harris, Harris, and
McLaughlin 1998;

\item {\it x-ray properties}: the excess in $S_N$ for giant cluster ellipticals 
over a value of $\sim4$ increases with the local galactic cluster
density, as determined from the x-ray temperature and position in the cluster
(West et al. 1995; Blakeslee et al. 1997).  
Straight correlations with x-ray temperature are also present, but
less obvious (Harris, Harris and McLaughlin 1998).

\end{itemize}

These cluster-wide correlations imply that GC systems in some giant
ellipticals and cD galaxies come from the cluster as a whole,
forming either in cluster gas before the individual galaxies
condensed, or forming in and around the individual galaxies but 
migrating to the cluster centers after getting stripped during
subsequent encounters.
Problems with stripping as a sole mechanism to increase $S_N$ in
cD galaxies were discussed above. The other option, whole cluster
formation, might account for the blue population because that formed
first in a cluster, and it tends to be at large radii around the
giant ellipticals, but the blue population is at most about half of the total
in giant ellipticals, so $S_N$ would not change much if the
blue GC's alone came from the galaxy cluster. 

McLaughlin et al. (1994) and Blakeslee and Tonry (1995) noted that only
the least dynamically
evolved clusters, with Bautz-Morgan types BM II and III, have cD
galaxies with high $S_N$ values. Giant, centralized elliptical galaxies
in the highly evolved BM I clusters have normal $S_N$. McLaughlin et al. (1994) suggested
that the more evolved BM I clusters have had time to dilute initially
high $S_N$ values in their centers with lower $S_N$ material from merged
and stripped neighbors. The less evolved BM II and III types are still
experiencing merging, as shown, for example, by multiple velocity
components in the nuclei of some galaxies (Blakeslee and Tonry 1992).
This implies that if cD's with large $S_N$ get their extra
GC's from ``partial'' stripping, then they have to do this quickly, before
they merge more completely with the remaining low-$S_N$ galaxies.
Blakeslee and Tonry (1995) also showed that only 20\% of the GC's in the Virgo
galaxies surrounding M87 would be required to account for the high $S_N$
in M87 if they were stripped, and that this same fraction is likely
to have been stripped for typical companion distances (based on Merritt
1988).

Other models for high $S_N$ in clusters involve GC formation in cooling
flows (Fabian, Nulsen, and Canizares 1984; Fall and Rees 1985). 
This idea is generally discounted now because
there are cD galaxies with low $S_N$ in clusters with high
x-ray emissions, as well as cD galaxies with high $S_N$ in 
clusters with low x-ray emission (West 1993; Harris, Pritchet and
McClure 1995; Kaisler et al. 1996).

Another model that accounts for high $S_N$ in the centers of
dense clusters, as well as the constancy in the globular cluster
luminosity function and characteristic mass, is the biased
GC formation model by West (1993). In this model, GC's and other
stars form only where the total density exceeds some specific value.
Since the background density is relatively high in the centers of clusters,
a lot of GC's formed there early on, making the blue GC population. 
The lower-density outskirts of clusters, and the lower-density clusters
and galaxy groups, formed relatively few GC because the critical
density was harder to reach.  There have been relatively few
comments on this model in the literature, 
but it is difficult to rule out considering
the popularity of biased galaxy formation in general (Kaiser 1984;
Davis et al. 1985; Bardeen et al. 1986).  However, there is
still some question 
about what is most constant, the number of GC per unit baryon mass
(McLaughlin 1999)
or the number of GC per unit total mass (including dark matter;
Blakeslee 1997).  The biasing model would seem to be most appropriate
in the latter case. 

An important observation that may shed some light on 
early star formation and merging in galaxy clusters is the
presence of planetary nebulae and diffuse star light 
between the galaxies.  Thuan and Kormendy (1977) found that
background light in the Coma cluster is nearly half (45\%) 
of the total light coming from the galaxies, and that the
color of this light is about the same blue as the outer halo of
the central cD, M87. This suggests that the cD halo and
the intergalactic stars were made from the same, presumably
stripped, material, and adds support to the idea, mentioned above,
that the extended GC system around M87 is from the cluster
at large (see also Weil et al. 1997).

Other evidence for intergalactic stars in clusters is the recent
observation of planetary nebula in Virgo (Arnaboldi et al. 1996;
Mendez et al. 1997) and Fornax (Theuns and Warren 1997), and the
observation of red giants in Virgo (Ferguson et al. 1998).
These studies also estimate that stray stars amount to an extra mass that is
about half the known galactic mass.

\section{A Constant Globular Cluster Formation Efficiency}

van den Bergh (1994) noted that globular clusters amount to about 2\%
of the total luminosities of the halo stars in both the Milky Way and
the LMC. This would imply a similar cluster formation efficiency in
two different galaxies. Other observations have suggested the
same thing. Richer et al. (1993) and 
Harris, Pritchet and McClure (1995) noted that $\sim1$\%
of the gravitating mass in clusters of galaxies turns into 
globular clusters.  Zepf and Ashman (1993) and Harris and Pudritz
(1994) similarly pointed out that the GC {\it mass} formation efficiency is
more constant than the {\it luminosity} efficiency, as measured by 
$S_N$.
Durrell et al. (1996) suggested that globular clusters represent
about 1\% of all star formation in order to get the metal
enrichment needed to account for the difference between the
GC metals and the associated galaxy metals. If the overall efficiency
of star formation in gas is $\sim10$\%, then the globular cluster
efficiency per unit {\it gas} mass would be about 0.1\%. 

Blakeslee, Tonry and Metzger (1997) used the velocity dispersions of
galaxy clusters to get the total cluster masses within 40 kpc
(including dark matter),
and found that the number of globular clusters per unit cluster mass
is about constant, equal to $\sim0.7$ GC's per $10^9$ M$_\odot$
of galaxy cluster. Considering an average GC mass of $10^6$ M$_\odot$, 
they converted this to $<0.1$\% (actually 0.0007) M$_\odot$ of 
globular clusters per $M_\odot$ of galaxy cluster mass. 

Harris, Harris and McLaughlin (1998) suggested that an average of $5.8$
globular clusters form in each $10^9$ M$_\odot$ of galaxy cluster baryonic 
mass,
so if the average GC mass is $3\times10^5$ M$_\odot$, which they
assumed, then the formation
efficiency is 0.0018 M$_\odot$ of GC per M$_\odot$ of total baryon mass. They
also suggested that the efficiency is generally {\it lower} in more {\it
massive} galaxy clusters, and that it is {\it larger} in {\it denser}
galaxy clusters, unless bright cluster ellipticals limit their own
brightnesses through star-formation induced winds (Blakeslee et al. 1997).

McLaughlin (1999) recovered the constant cluster formation efficiency
by considering the x-ray gas in clusters. He got a value of 0.0025
M$_\odot$ of GC for each M$_\odot$ of primordial gas in a
galaxy cluster.  This is about the same as the local mass efficiency
of bound cluster formation in the Milky Way (Harris and Pudritz 1994).

Observations of a constant efficiency for GC formation leave open a
question about the efficiency of non-clustered star formation (Larsen,
private communication).  If $S_N$ varies, then the number of GC per
non-clustered star varies, and if x-ray gas mass is needed to make the
efficiency of GC formation constant, then there must also be variations
in the ratio of GC mass to gas mass that does not form stars.  Why is
the GC formation efficiency more constant than the formation efficiency
of other stars? Do GC form first with this constant efficiency, and then
destroy the remaining clouds with a variable rate that depends on other
properties of the environment?

These studies show that globular clusters form in a more-or-less uniform
fashion in galaxy clusters, with an efficiency that is not much
different from that of normal open clusters. This probably implies that
the star formation processes were not much different in the early
Universe than they are today. How all of this relates to the bimodal
color distributions of globular clusters, to the likely stripping of
neighbors by giant cD galaxies, and to GC destruction over time, is not
clear. The complexity of this problem illustrates that globular clusters
today should be viewed in the context of the entire history of the
Universe, ranging from star formation and enrichment before most of the
galaxies formed, to galaxy evolution and globular cluster destruction
over time, to galaxy interactions, cluster stripping, and galaxy cluster
evolution.

\section{Summary}

The globular cluster specific frequency, $S_N$, is smaller by a factor
of $\sim6$ in spiral galaxies than in ellipticals (or a factor of
$\sim3$ if spiral galaxy fading and extinction are considered), and
smaller by another factor of $\sim5$ in ellipticals compared to giant cD
galaxies in Bautz-Morgan type II and III clusters. Some of this
variation is the result of changes in the mass-to-light ratio,
particularly among elliptical galaxies with increasing $L$, and some is
the result of changes in the total gas mass (seen in x-ray) to the
galaxy luminosity. Variations in $S_N$ by a factor of $\sim2$ seem
likely for merging galaxies, which make globular clusters during a
starburst phase and leave the system looking like an elliptical galaxy. 

The efficiency of globular cluster formation per unit total
mass, or per unit total {\it gas} mass, is remarkably constant
in different environments because most of the $S_N$ changes
are accounted for by changes in the ratio of total mass to 
luminosity.  Moreover, this near-constant mass efficiency is
about the same as the efficiency of bound cluster formation 
in galaxy disks today. This result implies that globular clusters
may have formed by normal star formation processes (Harris and Pudritz
1994; Elmegreen and Efremov 1997).   

Much of this globular cluster formation occurred very early in the
Universe, before the galaxies were pieced together. This gave the blue
population of globular clusters in elliptical galaxies and the halo
population in spirals (which have the lowest metal abundances of all).
The enrichment that resulted from this cluster formation, and from all
of the associated non-cluster star formation, provided the metals for
the galaxies that were soon to form in the same regions, from the
residual and recycled gas. 

When the elliptical galaxies formed in this second phase, more globular
clusters formed too, along with non-clustered stars, giving the red and
galaxy-correlated populations of globulars in the elliptical galaxies,
with their larger metal abundances. The analogous second phase of galaxy
building for spirals occurred primarily in the disks, where additional,
metal-rich globular clusters formed too.

After the globular clusters and galaxies formed, the galaxies began to
interact, first stripping off the peripheral clusters that were most
loosely bound, and collecting these, along with loosened stars, in the
galaxy cluster potential wells. The interactions continued
by direct merging of gassy
systems in the field. This phase formed the cD's in clusters of
galaxies, with their relatively large $S_N$ values compared to cluster
ellipticals, and it also formed some of the ellipticals in the field,
with their relatively low $S_N$ values compared to cluster ellipticals. 

While this scenario makes a lot of sense, and goes a long way toward
explaining $S_N$ and other properties of globular cluster systems
and their associated galaxies, there are some perplexing issues that
have no satisfactory explanations at all.  We list them here for future
study:

\begin{itemize} \item Why is the globular cluster luminosity function,
or mass function, so constant from place to place? Does this constancy
imply that it is the same as the initial cluster mass function? If so,
then why is there a characteristic mass (when all other ``normal'' star
formation operates in a scale-free interstellar medium). If today's
cluster mass function is not the initial function, then how could
cluster erosion make today's function the same in all environments?

\item Why is the bound cluster formation efficiency so constant from
place to place and over time? How much does it vary with total star
formation rate (Larsen and Richtler 1999) or other local properties of
the environment? If $S_N$ varies even when the efficiency of globular
cluster formation
is constant, does this mean that the efficiency of non-clustered
star formation is varying too (S. Larsen, private communication)? Why
would the efficiency of GC formation be more constant than the
efficiency of all star formation?

\item When and how did the red population of globular clusters form
in bright elliptical galaxies? Did it form during interactions or
{\it in situ}?  Did any other sub-populations of globular
clusters form during interactions or at later times and remain 
obscure because of a conspiracy between age and metallicity 
(Kissler-Patig et al. 1998b)?  Did the blue population of
globular clusters in giant ellipticals 
come from stripped neighbors? If so, why
did it get so well distributed throughout the galaxy and not
just remain in the outer regions?

\item Why does some star formation produce bound clusters and other star
formation not? That is, why does the local efficiency of star formation
in a cloud core vary (being higher where bound clusters form)? Is the
high efficiency the result of compression and triggering, as appears to
be the case for many galactic disk clusters today (Elmegreen et al.
1999)? Do the high pressures required for globular cluster formation
(Elmegreen and Efremov 1997) also imply some type of triggering, or
cloud collisions (Kumai et al. 1993b). If so, then why do globular
clusters also form in very low mass galaxies, where the velocity
dispersion and total gas pressure are extremely low (van den Bergh
1995b)?

Some of these problems are related to star formation in general, and
others to galaxy formation and evolution. We might have to understand
all of these aspects of globular clusters together before we can
understand any one part of the picture by itself. 

\end{itemize}

\acknowledgements

Financial support for attending this conference was kindly provided
by the Anglo American Chairman's Fund, and SASOL.  Helpful comments
on the manuscript by Drs. T. Richtler, D. McLaughlin, and S. Larsen
are appreciated.

\end{article}
\end{document}